\documentclass[pra,10pt,letterpaper,draft,bibnotes,notitlepage,final,balancelastpage]{revtex4}
\usepackage{epsfig,graphicx,graphics,times,amssymb,amsmath,amsfonts,bm,bbm}

\begin{document}

\title{Investigation of quantum entanglement simulation by random variables theories augmented by
either classical communication or nonlocal effects}
\author{Akbar Fahmi }
\email{ fahmi@theory.ipm.ac.ir}

\affiliation{Department of Philosophy of Science, Sharif University of Technology, Tehran,14588, Iran\\
School of Physics, Institute for Research in Fundamental Sciences (IPM), P.O.Box 19395-5531, Tehran, Iran}
\date{{\small{\today}}}

\begin{abstract}
Bell's theorem states that quantum mechanics is not a locally causal theory. This state is often interpreted as nonlocality in quantum mechanics. Toner and Bacon [Phys. Rev. Lett. \textbf{91}, 187904 (2003)] have shown that a shared random-variables theory augmented by one bit of classical communication exactly simulates the Bell correlation in a singlet state. In this paper, we show that in Toner and Bacon protocol, one of the parties (Bob) can deduce another party's (Alice) measurement outputs, if she only informs Bob of one of her own outputs.
Afterwards, we suggest a nonlocal version of Toner and Bacon protocol wherein classical communications is replaced by nonlocal effects, so that Alice's measurements cause instantaneous effects on Bob's outputs. In the nonlocal version of Toner and Bacon's protocol, we get the same result again. We also demonstrate that the same approach is applicable to Svozil's protocol.
\end{abstract}
\maketitle

\section{Introduction}

The development of quantum mechanics (QM) in the early twentieth
century obliged physicists to radically change some of the
concepts they employed to describe the world. Entanglement was first viewed as a source of some paradoxes,
most noticeably the Einstein-Podolsky-Rosen paradox (EPR)
\cite{EPR}, which explicitly states that any physical theory
must satisfy both local and realistic conditions. These conditions then manifest
themselves in the so-called Bell inequality \cite{Bell,Bell1}. However, this inequality is violated by quantum predictions. This violation is often
referred to as quantum nonlocality and has been recognized as the
most intriguing quantum feature.
The Bell inequality has been derived in different ways
\cite{CHSH,CH} and over the past 30 years, various types of Bell's inequalities have undergone a wide variety of experimental tests. All of them demonstrate strong indications against local hidden variable theories \cite{As}. These results are often interpreted as nonlocality in quantum mechanics.

Now we face with an interesting question: \emph{How much nonlocality or classical resources are required to simulate quantum systems?} An insightful approach for such simulation is to characterize information processing tasks in which two parties share random classical resources and communicate various types of classical bits. In this direction, simulation of Bell's correlation by shared random-variable (SRV) models augmented by classical communication or nonlocal effects has recently attracted a lot of attention \cite{Mau,Brass,St,Bacon,De,Non,svozil,Gisin1,Cerf1,Cav1,Cav2}. The question of whether a simulation can be done with a finite
amount of communication has been considered independently by Maudlin \cite{Mau}, Brassard, Cleve, and Tapp \cite{Brass},
and Steiner \cite{St}. Brassard, Cleve, and Tapp showed that $8$ bits of communication suffice for a
perfect (analytic) simulation of the quantum predictions. Steiner,
followed by Gisin and Gisin \cite{Gisin1}, showed that if one allows the number of
bits to vary from one instance to another, then $2$ bits are sufficient
on average. It also has been shown that if many singlets have to be
simulated in parallel, then block coding could be used to reduce the
number of communicated bits to $1.19$ bits on average
\cite{Cerf1}. A few years later, Toner and Bacon \cite{Bacon}
improved these results and showed that a simulation of
the Bell correlation (singlet state) is possible by implementing only one
bit of classical communication. Toner and Bacon concluded that their results prove minimal amount, one bit, is sufficient to simulate projective
measurements on Bell states. In the same way, Svozil suggested another model
\cite{svozil} which is based on the Toner and Bacon protocol (TB protocol) and is more nonlocal. Afterward, Tessier \emph{et al.} have
shown that it is possible to reproduce the quantum-mechanical
measurement predictions for the set of all $n$-fold products of
Pauli operators on an $n$-qubit GHZ state using only Mermin-type random variables and $n-2$ bits of classical communication \cite{Cav1}. With a similar approach, Barrett \emph{et al.} proposed a communication-assisted random-variables model
that yields correct outcome for the measurement of any
product of Pauli operators on an arbitrary graph state \cite{Cav2}.

Independently of the above developments, Popescu and Rohrlich \cite{PR1}
have dealt with a question: Can there be stronger correlations
than the quantum mechanical correlations that remain causal (not allow signaling)? Their answer draws upon exhibiting an
abstract nonlocal box wherein instantaneous communication remains
impossible. This nonlocal box is such that the Clauser-Horne-Shimony-Holt (CHSH)
inequality is violated by the algebraic maximum value
of $4$, while quantum correlations achieve at most $2\sqrt{2}$ \cite{PR1,JM}. There is a question of interest: If perfect nonlocal boxes would not violate causality, why do the laws of quantum mechanics only allow us to implement nonlocal boxes better than anything classically possible, yet not perfectly \cite{Bra2}?
Recently, van Dam and Cleve considered communication complexity as
a physical principle to distinguish physical theories from nonphysical ones.
They proved that the availability of perfect nonlocal boxes makes the
communication complexity of all Boolean functions trivial \cite{Dam}. Afterwards,
Brassard \emph{et al.} \cite{Bra2} showed that in
any world in which communication complexity is nontrivial, there should be a bound
on how much nature can be nonlocal. Besides, Pawlowski \emph{et al.} \cite{IC} defined information
causality as a candidate for one of the fundamental assumptions of quantum
theory which distinguishes physical theories from nonphysical ones. In fact, Svozil's model has simulated the nonlocal Box as suggested in \cite{PR1}.

In this paper, we review the TB model which simulates Bell correlations \cite{Bacon}. We show that if Alice informs Bob from one of her outputs, he can deduce Alice's measurement results with no need for more classical communications or other resources. Afterwards, we propose a nonlocal version of the TB protocol (NTB) and Svozil protocol (NS), in order to construct a similar structure as the nonlocal Box model \cite{PR1}. The NTB (NS) model is an imaginary device (includes two input-output ports, one at Alice's location and another at Bob's location), in which classical communications are replaced with instantaneous nonlocal effects. In the NTB (NS) model, we get the same result as previous ones. Moreover, it can be proved that the availability of a perfect NTB protocol makes the communication complexity of all Boolean functions trivial.

This article is organized as follows: In Sec. II we briefly review the original TB protocol \cite{Bacon} and show that if Alice only informs Bob from one of her outputs, he can infer Alice's outputs without any need for more classical communications. Moreover, we apply our approach to Svozil's protocol \cite{svozil}. In Sec. III we extend the TB protocol to a nonlocal case by replacing classical communication bits (cbit) with nonlocal effects. In this new protocol, Alice's measurements cause a nonlocal effect in Bob's outputs. We also show that in this situation, if Alice only informs Bob from one of her outputs, he can deduce Alice's measurement outputs without any need for more classical communications. In Sec. IV we summarize our results.

\section{Bob infers Alice's Measurement outputs in the TB and The Svozil protocols}

In this section, we briefly review the TB and Svozil protocols and show how in these protocols
Bob can infer Alice's measurement outputs.

\subsection{Toner and Bacon protocol}
Consider Bell's experiment setup in which a source emits two spin half ($\frac{1}{2}$) particles (or qubits) to two spatially separate parties (conventionally named Alice and Bob). The states of shared qubits is the entangled Bell singlet state (also known as an EPR pair) $|\psi^{-}\rangle={1 \over \sqrt{2}} \left(|+ \rangle_A |- \rangle_B - |- \rangle_A |+ \rangle_B\right)$. The spin states $|+\rangle$, $|-\rangle$ are
defined with respect to a local set of coordinate axes:
$|+\rangle$ ($|- \rangle$) corresponds to spin-up (spin-down)
along the local $\hat{z}$ direction. Alice and
Bob each measure their qubit's spin along a direction parametrized
by three-dimensional unit vectors $\hat{a}$ and $\hat{b}$, respectively. Alice and Bob obtain results, $\alpha\in\{+1,-1\}$ and $\beta \in\{+1,-1\}$, respectively, which
indicate whether the spin was pointing along ($+1$) or opposite ($-1$)
the direction each party chose to measure. Alice and Bob's marginal outputs appear random, with expectation values $\langle \alpha\rangle = \langle \beta \rangle =0$; joint expectation values are
correlated such that $\langle \alpha \beta \rangle = - \hat{a}\cdot \hat{b}$.

Bell's correlations have three simple properties: (i) if $\hat a= \hat b$, then Alice and Bob's outputs are perfectly anticorrelated, i.e., $\alpha = - \beta$; (ii) if Alice (Bob)
reverses her (his) measurement axis $\hat a\rightarrow -\hat a$ ($\hat b\rightarrow -\hat b$), the outputs are flipped  $\alpha\rightarrow -\alpha$ ($\beta\rightarrow -\beta$); and (iii)
the joint expectation values are only dependent on $\hat a$ and $\hat b$ via
the combination $-\hat a \cdot \hat b$. Now in order to answer the aforementioned question about what classical resources are required to simulate Bell states correlations, Toner and Bacon revised the original Bell's model \cite{Bell} and obtained a minimal required number of
classical resources which need to simulate the Bell states. They gave a local hidden variables model augmented by just one bit of classical communication to reproduce these three properties for all possible axes [Bell's model fails to reproduce property (iii)].

In the TB protocol, Alice and Bob share two independent random variables $\hat{\lambda}_{1}$ and $\hat{\lambda}_{2}$ which are real three-dimensional unit vectors. They are independently chosen and uniformly distributed over the unit sphere (infinite communication at this stage). Alice measures along $\hat{a}$; Bob measures along $\hat{b}$. They obtain $\alpha\in\{+1,-1\}$ and $\beta\in\{+1,-1\}$, respectively. The TB protocol proceeds as follows:

(1.) Alice's outputs are $\alpha=-\rm sgn(\hat{a}\cdot\hat{\lambda}_{1})$.

(2.) Alice sends a single bit $ c \in \{-1, +1\}$ to Bob where
$c=\rm sgn(\hat{a}\cdot\hat{\lambda}_{1})\rm sgn(\hat{a}\cdot\hat{\lambda}_{2})$.

(3.) Bob's outputs are
$\beta=\rm sgn[\hat{b}\cdot(\hat{\lambda}_{1}+c\hat{\lambda}_{2})]$,
where the $\rm sgn$ function is defined by $\rm sgn(x)=+1$ if $x \geq 0$ and
$\rm sgn(x)=-1$ if $x < 0$.

The joint expectation value
$\langle\alpha\beta\rangle$ is given by:
\begin{eqnarray*}
\langle \alpha \beta \rangle = E\Bigg\{-\rm sgn (\hat{a} \cdot \hat{\lambda}_{1}) \times
\sum_{d=\pm 1} \frac{(1+cd)}{2}\rm sgn \left[\hat b
\cdot(\hat{\lambda}_{1} + d \lambda_{2})\right] \Bigg\}
\end{eqnarray*}
where $E\left\{ x \right\} ={1 \over (4\pi)^2} \int d \hat{\lambda}_{1} \int d
\lambda_{2}\, x$, $c = \rm sgn (\hat{a} \cdot \hat{\lambda}_{1})\, \rm sgn (\hat{a} \cdot\hat{\lambda}_{2}
)$. This integral can be evaluated which gives $\langle \alpha \beta \rangle
= - \hat a \cdot \hat b$, as required.


\emph{Remark 1.}-- In this article, we use the terminology in which the parties have complete control over the shared random-variables without referring to each other \cite{De,Cerf1,svozil,Cerf,BGS}. Therefore, we use SRV and hidden random-variables (HRV) interchangeability.

\emph{Remark 2.}-- TB claimed that Bob obtains ``no information" about Alice's outputs from the cbits communications. In the next subsection, we show that it is not correct.

\begin{figure}
\centering
\includegraphics[height=6cm,width=7.5cm]{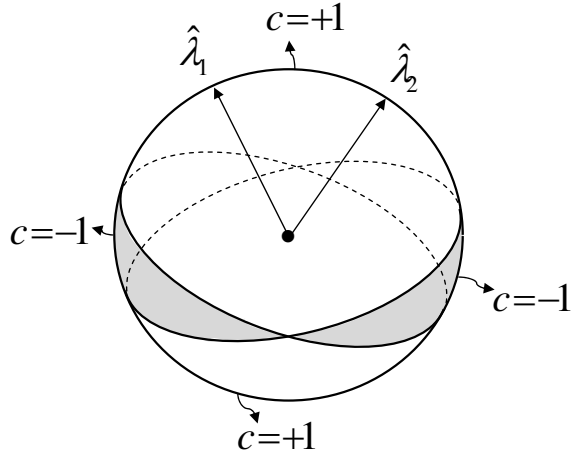}
\caption{The random unit vectors $\hat{\lambda}_{1}$ and $\hat{\lambda}_{2}$ divide the Poincare´ sphere into four parts. If Alice's measurement setting $\hat{a}$ lies in the shaded region, she sends $c=-1$ and if her measurement axis lies in the unshaded region, she sends $c=+1$ to Bob. Toner and Bacon deduce that Bob obtains no information about Alice's output from the communication.
}\label{TB}
\end{figure}

\subsection{Bob Finds Alice's measurement output in the Toner and Bacon protocol}
In this subsection, we shall show that in the TB protocol Bob can deduce Alice's measurement output, if she just notifies Bob of one of her outputs without using other classical communications. At the first stage, let us define some useful quantities. We define unit vectors $\hat{\lambda}_{1}$ and $\hat{\lambda}_{2}$ in the spherical coordinate
($\theta,\phi$), at the ranges of $\theta\in(0,\pi)$ and
$\phi\in(0,2\pi)$ and dividing $\theta$ and $\phi$ into $N$ equal parts, so that
$\pi/N=\delta\ll1$ as $N\rightarrow\infty$.

Now, we consider a subset of SRV $\hat{\lambda}_{1}(\theta_{1},\phi_{1}), \hat{\lambda}_{2}(\theta_{2},\phi_{2})$ in the $xy$ plane which are represented by $\left\{\hat{\lambda}^{xy}_{1}(\theta_{1}=\pi/2,\phi_{1}=l\delta),\hat{\lambda}^{xy}_{2}(\theta_{2}=\pi/2,\phi_{2}=k\delta)\right\}$,
where $l,k=0,1,...,N$. For simplicity, we do not refer to
$\theta_{1,2}=\pi/2$ and denote them as
$\hat{\lambda}^{xy}_{i}(\theta_{i}=\pi/2,\phi_{i}=t\delta)\equiv\hat{\lambda}^{xy}_{i,t}$, ($i=1,2$, and $t=l,k$). Here, $\hat{\lambda}^{xy}_{i,t}$ ($i=1,2$) means that the SRV $\hat{\lambda}^{xy}_{i,t}$ makes the azimuthal angle $t\delta$ with the $\hat{x}$ axis. We select a specific subset of SRV and consider the collection:
\begin{eqnarray}\label{xy}
\left\{\left(\hat{\lambda}^{xy}_{1,l},\hat{\lambda}^{xy}_{2,l+1}\right)\right\},
\end{eqnarray}
where, $\hat{\lambda}_{1,l}^{xy}\cdot\hat{\lambda}_{2,l+1}^{xy}=\cos\delta$, $\hspace{.1cm}\delta=\frac{\pi}{N}\ll1$, and random vectors $\hat{\lambda}_{i,l+1}^{xy}$ ($i=1,2,\hspace{.2cm}l=0,...,N$) are given by applying rotation operators $R(\hat{z},\delta)\in$ SO(3) (around the $\hat{z}$ axis) on $\hat{\lambda}_{i,l}^{xy}$ $(\hat{\lambda}_{i,l+1}^{xy}=R(\hat{z},\delta)\hat{\lambda}_{i,l}^{xy},\hspace{.1cm}\forall l)$.

\begin{figure}
\centering
\includegraphics[height=6.2cm,width=6.3cm]{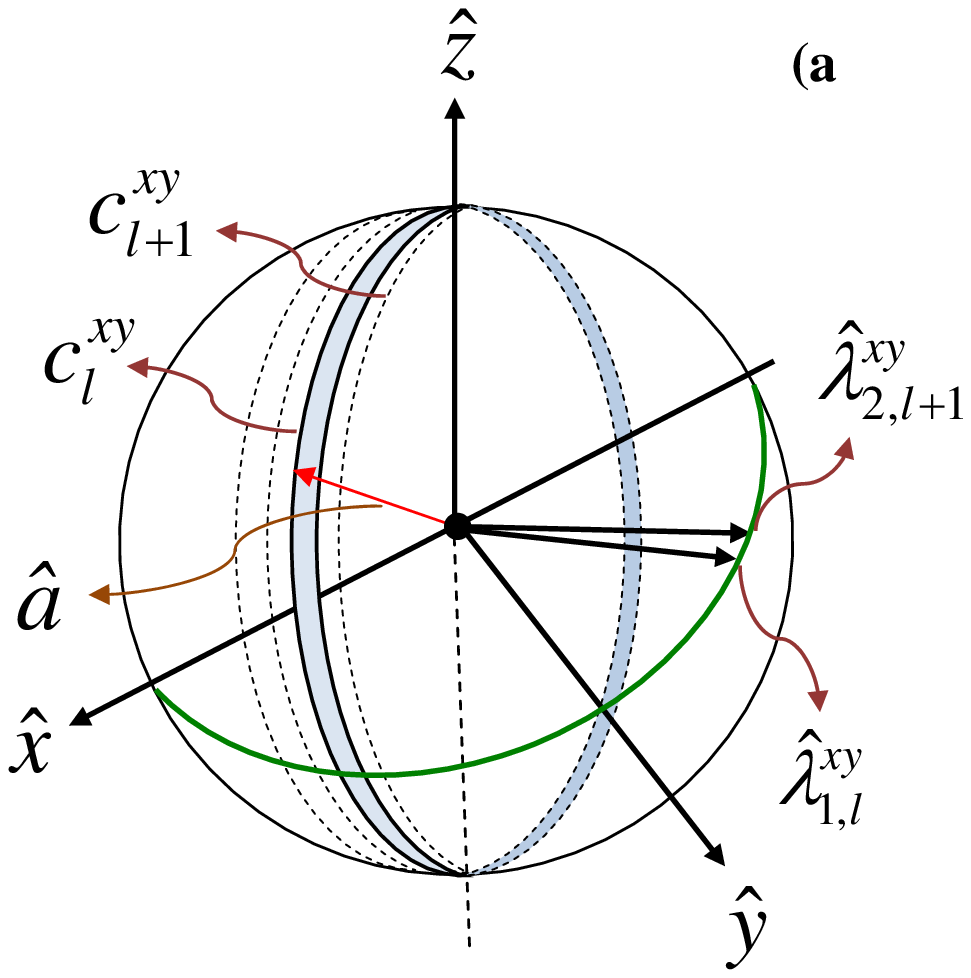}
\includegraphics[height=6.6cm,width=5.6cm]{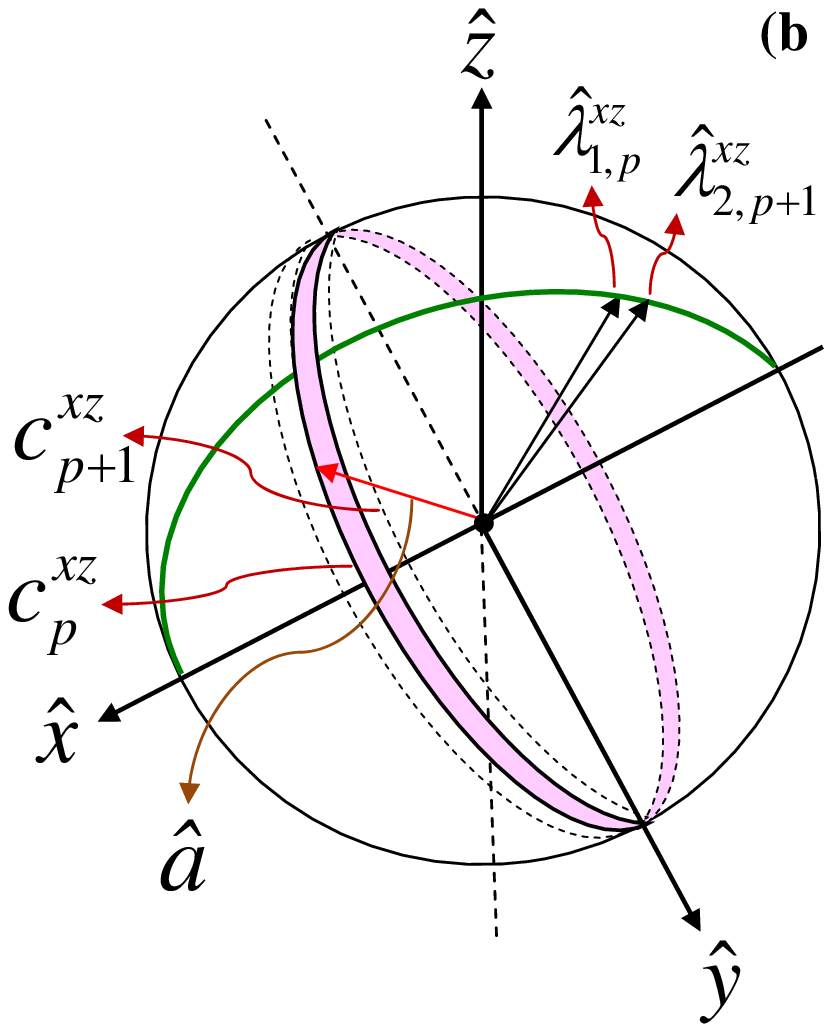}
\includegraphics[height=6.8cm,width=6cm]{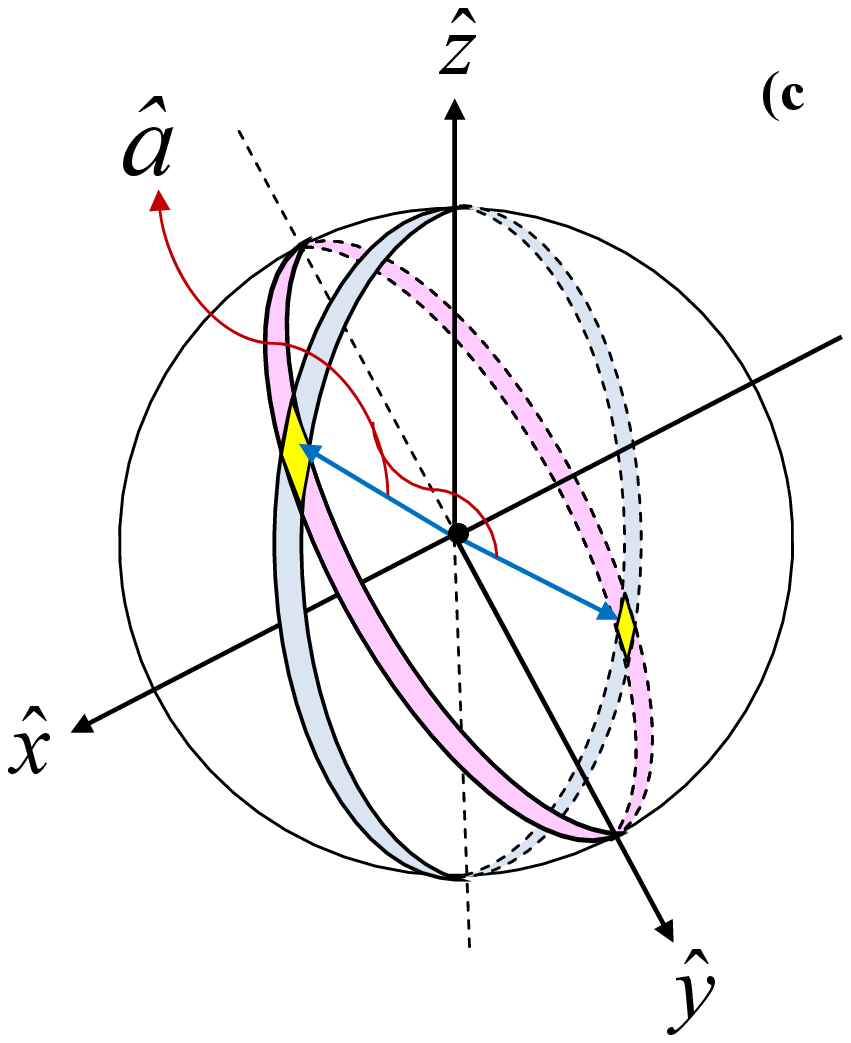}
\caption{(Color online). (a) Subsets of shared random variables lie in the $xy$
plan. The blue zone corresponds to random variables
$(\hat{\lambda}_{1,l}^{xy},\hat{\lambda}_{2,l+1}^{xy})$ with
$c_{l}^{xy}=-1$ (Fig.\ref{TB}). According to the $c$ definition, Alice's
measurement setting $\hat{a}$ must lie in a plane with unit
vector $\hat{\lambda}_{1,l}^{xy}$ (blue zone in the $l$-th round
of experiment). The azimuthal angle ($\hat{a}$) is equal to
$l\delta\pm\pi/2$. (b) Subsets of shared random variables lie in
the $xz$ plane. The red zone corresponds to one set of random
variables $\hat{\lambda}_{1,p}^{xz},\hat{\lambda}_{2,p+1}^{xz}$
with $c_{p}^{xz}=-1$ (Fig.\ref{TB}). Alice's measurement setting
$\hat{a}$ must lie in a plane with unit vector
$\hat{\lambda}_{1,p}^{xz}$ (red zone in the $p$-th round of experiment). In fact, the thin strips sweep the surface of the unit sphere
so that $c=-1$ for $l$-th and $p$-th strips and $c=+1$ elsewhere.
(c) These two strips cross each other at two small areas and
consequently, Alice's measurement setting $\hat{a}$ is in the same (or in the opposite) direction of the unit vector which connects the origin of the Poincare´ sphere to the cross points.}\label{1}
\end{figure}

The sequences of communicating classical bits corresponding to the above set of random variables are represented by $c^{xy}_{k}(\hat{a},\hat{\lambda}_{1,k}^{xy},\hat{\lambda}_{2,k+1}^{xy})$. With due attention to the SRV subset, the sign of one of the communicating classical bits will switch to a negative value as is shown by the following statements:
\begin{eqnarray}\label{C}
...,c^{xy}_{l-2}(\hat{a},\hat{\lambda}_{1,l-2}^{xy},\hat{\lambda}_{2,l-1}^{xy})&=&+1,\nonumber\\
c^{xy}_{l-1}(\hat{a},\hat{\lambda}_{1,l-1}^{xy},\hat{\lambda}_{2,l}^{xy})&=&+1,\nonumber\\
c^{xy}_{l}(\hat{a},\hat{\lambda}_{1,l}^{xy},\hat{\lambda}_{2,l+1}^{xy})&=&-1,\nonumber\\
c^{xy}_{l+1}(\hat{a},\hat{\lambda}_{1,l+1}^{xy},\hat{\lambda}_{2,l+2}^{xy})&=&+1,....
\end{eqnarray}
In the above relations, we assumed that in the $l$-th round of the protocol, the sign of the communicated bit has been changed. According to this sequence, Bob deduces that Alice's measurement
setting lies in a plane with the unit vector
$\hat{\lambda}_{1,l}^{xy}$ [$l$-th strip Figs. \ref{TB} and
\ref{1}(a)]. Therefore, in the spherical coordinate, the azimuthal angle of
$\hat{a}$ ($\phi_{\hat{a}}$) is equal to $l\delta\pm\pi/2$, with
the uncertainty factor $\delta$. One should notice that in this situation, Bob cannot yet fix the
polar angle ($\theta_{\hat{a}}$).

In the second stage, in order to find $\theta_{\hat{a}}$, we can
select the random variables $\hat{\lambda}_{1}$ and
$\hat{\lambda}_{2}$ in a plane with unit vector
$\hat{\lambda}_{1,l}^{x'y'}$. It can be obtained by rotating the $\hat{x}$
axis by amounts $\phi=l\delta+\pi/2$ or $\phi=l\delta-\pi/2$ around the
$\hat{z}$ axis. These hidden variables are obtained by
$\left\{\hat{\lambda}^{x'z}_{1}(\theta,\phi=l\delta\pm\pi/2),\hat{\lambda}^{x'z}_{2}(\theta,\phi=l\delta\pm\pi/2)\right\}$
in the Poincare´ sphere coordinates. However, similar to the first stage, we consider $\hat{\lambda}_{1}$ and $\hat{\lambda}_{2}$ in the $xz$ plane which are represented by $\left\{\hat{\lambda}^{xz}_{1}(\theta_{1}=p\delta,\phi_{1}=0),\hat{\lambda}^{xz}_{2}(\theta_{2}=q\delta,\phi_{2}=0)\right\}
\equiv\left\{\hat{\lambda}^{xz}_{1,p},\hat{\lambda}^{xz}_{2,q}\right\}$,
where $p,q=0,1,...,N$. Here, $\hat{\lambda}^{xz}_{i,t}$ ($i=1,2$) means that the SRV $\hat{\lambda}^{xz}_{i,t}$ makes polar angle $t\delta$ with the $\hat{z}$ axis. We select a specific subset of SRV and consider a collection similar to Eq. (\ref{xy}):
\begin{eqnarray}\label{xz}
\left\{\left(\hat{\lambda}^{xz}_{1,p},\hat{\lambda}^{xz}_{2,p+1}\right)\right\},
\end{eqnarray}
where, $\hat{\lambda}_{1,p}^{xz}\cdot\hat{\lambda}_{2,p+1}^{xz}=\cos\delta$, $\delta=\frac{\pi}{N}\ll1$, and other random vectors such as $\left(\hat{\lambda}^{xz}_{1,p+1},\hat{\lambda}^{xz}_{2,p+2}\right)$ ($i=1,2,\hspace{.2cm}p=0,...,N$) are given by applying rotation operators $R(\hat{y},\delta)\in$ SO(3) on the $\left(\hat{\lambda}^{xz}_{1,p},\hat{\lambda}^{xz}_{2,p+1}\right)$, $\hspace{.1cm}\hat{\lambda}_{i,p+1}^{xz}=
R(\hat{y},\delta)\hat{\lambda}_{i,p}^{xz},\hspace{.1cm}\forall p$. The corresponding sequences of communicating cbits are given by similar relations to (\ref{C}) with
$c^{xy}_{l}(\hat{a},\hat{\lambda}_{1,l}^{xy},\hat{\lambda}_{2,l+1}^{xy}) \rightarrow c^{xz}_{p}(\hat{a},\hat{\lambda}_{1,p}^{xz},\hat{\lambda}_{2,p+1}^{xz})$.
According to this sequence, Bob deduces that $\hat{a}$ lies in a plane with unit vector
$\hat{\lambda}_{1,p}^{xz}$ [in the $p$-th strip, Figs. \ref{TB}
and \ref{1}(b)], with the uncertainty factor $\delta$.

Here, we have restricted selected shared random variables to two special subsets (\ref{xy}) and (\ref{xz}) because they are
sufficient for Bob to deduce Alice's measurement outputs. 
These two strips encounter each other at two points. Alice's
measurement setting $\hat{a}$ is in the same (or in the opposite) direction of the unit vector that connects the origin of the Poincare´ sphere to the
crossover points [Fig. \ref{1}(c)].

Now, if the parties collaborated and selected a specific random variables, for example, ${\lambda}_{r}$, and Alice informs Bob of only one of her outputs, Bob can deduce Alice's
measurement setting without any need for further information. For
example, if Alice sends
$\alpha_{r}=-sgn(\hat{a}.\hat{\lambda}_{r})=+1$ to Bob, he can
deduce the $\hat{a}$ direction lies in the up (down) semicircle.


Our approach is not restricted to the above selected subsets of hidden variables, The parties can use
other sets of SRV such as $\hat{\lambda}^{yz}_{1}$ and
$\hat{\lambda}^{yz}_{2}$ which belong to the $yz$ plane to get the same results.

\subsection{Svozil's protocol}
Svozil has suggested a new type of shared random-variable theory augmented by one bit of classical communication which is stronger than quantum correlations \cite{svozil}. It violates the CHSH inequality by $4$, as compared to the quantum Tsirelson bound $2\sqrt{2}$. Svozil's protocol is similar to the Toner and Bacon protocol \cite{Bacon}, but just requires only a single random variable ${\lambda }$. The another random variable ${\Delta} (\omega )$ is obtained by rotating ${\hat \lambda }$ clockwise around the origin by angle $\omega$ with a constant shift for all experiments ${\Delta} (\omega )=\lambda +\omega$. Alice's outputs are given by $\alpha  =- {\rm sgn}({\hat a} \cdot {\hat \lambda }) =-{\rm sgn}\left[\cos ({a} - { \lambda } )\right]$ and she sends classical bits $c(\omega) = {\rm sgn}({\hat a} \cdot {\hat \lambda } ){\rm sgn}\left[{\hat a} \cdot {\hat \Delta} (\omega)\right]= {\rm sgn}\left[\cos ({ a} - { \lambda } )\right] {\rm sgn}\cos \left[{ a} - { \Delta} (\omega)\right]$ to Bob. Bob's outputs are given by $\beta (\omega )=  {\rm sgn}[{\hat b} \cdot ({\hat \lambda } +c(\omega){\hat \Delta} (\omega))]$. If we let $\omega$ changes randomly on the Poincare´ sphere, then Svozil's protocol becomes the TB protocol (with uniform distribution). In the general case $0\le \omega \le \pi /2$, the correlation function is given by
\begin{eqnarray*}
E(\theta ,\omega )= \left\{
\begin{array}{ll}
-1 & \text{ for } \;\; 0\le \theta \le {\omega \over 2} ,  \\
-1 +{2\over \pi}(\theta -{\omega \over 2} )&\text{ for } \;\; {\omega \over 2} < \theta \le {1 \over 2}(\pi - \omega) ,   \\
-2(1-{2 \over \pi } \theta ) &\text{ for } \; \; {1 \over 2}(\pi - \omega) < \theta \le {1 \over 2}(\pi   + \omega ) , \\
1+ {2\over \pi }(\theta-\pi +{\omega \over 2} ) &\text{ for } \;\; {1 \over 2}(\pi   + \omega )  < \theta \le \pi - {\omega \over 2} , \\
1 & \text{ for } \;\; \pi - {\omega \over 2} < \theta \le \pi .
\end{array}        \right.
\label{e-2004-brainteaser-2}
\end{eqnarray*}
The correlation function $E(\theta ,\omega )$ is stronger than quantum correlations for all nonzero values of $\omega$. The strongest correlation function is obtained for $\omega =\pi /2$, where the two random-variable directions ${ \lambda }$ and ${ \Delta} ={ \lambda }+ \pi /2$ are orthogonal and the information of classical bits $c (\pi /2)$ are about the location of ${ a}$
within two opposite quadrants. In the case of $\omega =\pi /2$, the CHSH inequality $\vert E({ a} ,{ b} )- E({ a} ,{ b} ' )\vert+ \vert E({ a}' ,{ b} )- E({a} ',{ b} ') \vert \le 2$ is violated by the maximal algebraic value of $4$, for ${a} =0 $, ${a}'=\pi /2$, ${b} =\pi /4$, ${b}'= 3\pi /4$ (which are the same as the largest possible value which has been suggested by Popescu and Rohrlich's nonlocal box model \cite{PR1}). For $\omega =0$, the classical linear correlation function $E(\theta )=  2\theta /\pi -1$ is recovered, as could be expected.


\subsection{Bob finds Alice's measurement outputs in Svozil's protocol}

In this subsection, we consider Svozil's \cite{svozil} arguments and show that Bob can deduce Alice's measurement setting and outputs by using cbits and only one of the Alice outputs, without asking for any further information at the end of the protocol. Here, we select the $\omega=\frac{\pi}{2}$ case and consider the subset of SRV as:
\begin{eqnarray}\label{}
\left\{\left(\hat{\lambda}(k\delta),\hat{\Delta}(k\delta)\right)\right\},
\end{eqnarray}
where, $\delta=\pi/N\ll1$, $N\rightarrow\infty$, and $k=0,...,N$, and other random vectors are related to
each other by rotating $\hat{\lambda}_{k}(\equiv\hat{\lambda}(k\delta))$ and $\hat{\Delta}_{k}(\equiv\hat{\Delta}(k\delta))$ clockwise around the origin by an angle $\delta$. It acts as a constant shift for all experiments, i.e., $\lambda_{k+1}=\lambda_{k}+\delta$.

\begin{figure}
\centering
\includegraphics[height=8.5cm,width=8cm]{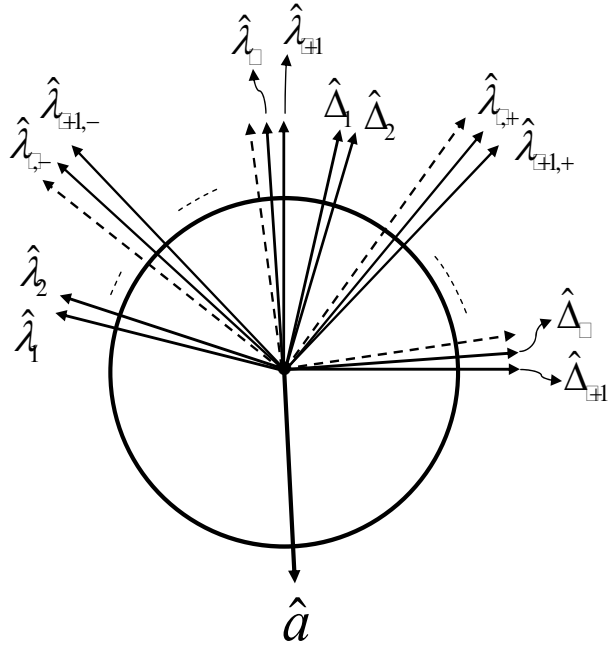}
\caption{
Svozil's protocol for the case of $\omega=\frac{\pi}{2}$.
Here $\hat{\lambda}_{k}$ is given by
rotating $\hat{\lambda}_{k-1}$ $(\forall k)$ clockwise around the origin
by $\delta$, and $\hat{\lambda}_{k,\pm}=\hat{\lambda}_{k}\pm\hat{\Delta}_{k}$.
According to the $c_{k}={\rm sgn}(\hat{a}\cdot\hat{\lambda}_{k}){\rm sgn}(\hat{a}\cdot\hat{\Delta}_{k})$ definition, in the $l+1$-th
round of the protocol the sign of the communicated bits change and consequently $\hat{a}$ becomes parallel (or
antiparallel) to $\hat{\lambda}_{l}$.}
\label{3}
\end{figure}

The sequences of communicating classical bits corresponding to the above set of random variables are represented by $c_{k}(\pi/2, \hat{\lambda}_{k})$.
With due attention to the SRV subset, after some of the communicating classical bits, the sign of $c_{k}$s will switch to opposite values as are given by the following quantities:
\begin{eqnarray*}
...,c_{l-1}(\pi/2, \hat{\lambda}_{l-1})=+1, c_{l}(\pi/2,
\hat{\lambda}_{l})=+1,\\ c_{l+1}(\pi/2, \hat{\lambda}_{l+1})=-1,
c_{l+2}(\pi/2, \hat{\lambda}_{l+2})=-1,...,
\end{eqnarray*}
where, in the $l+1$-th round of the protocol the
sign of communicated bits has changed (Fig. \ref{3}).

\emph{Remark 3.}-- With note to $c_{k}$ definition and selected
random variables, if $\hat{a}$ lies in $(+\hat{\lambda}_{k},+\hat{\Delta}_{k})$ or $(-\hat{\lambda}_{k},-\hat{\Delta}_{k})$ intervals, $c_{k}=+1$ and for other ranges $c_{k}=-1$:
\begin{eqnarray*}
&&\hat{a}\in(+\hat{\lambda}_{k},+\hat{\Delta}_{k})\vee(-\hat{\lambda}_{k},-\hat{\Delta}_{k})\rightarrow
c_{k}=+1,\\
&&\hat{a}\in(+\hat{\lambda}_{k},-\hat{\Delta}_{k})\vee(-\hat{\lambda}_{k},+\hat{\Delta}_{k})\rightarrow
c_{k}=-1,\hspace{.1cm}\forall k.
\end{eqnarray*}

As stated by the above sequence, Bob deduces that $\hat{a}$ is in the same (or in the opposite) direction as the unit vector $\hat{\lambda}_{l}$, with the uncertainty factor $\delta$.
At this stage, if Alice only informs Bob of one of her outputs, for example,
$\alpha_{r}(\hat{a},\lambda_{r})=- {\rm sgn}(\hat{a}.\hat{\lambda}_{r})=+1$
(or $-1$), Bob deduces that the $\hat{a}$ direction lies in the down
(up) semicircle. Thus, he deduces Alice's outputs and measurement
setting, without any need for further information \cite{Note1}.

\section{Bob infers Alice's Measurement outputs in SRV theories augmented by NonLocal Effects}

In the above approaches, Bob only uses a few parts of classical communication bits, without referring to his measurement's outputs. This leads us to ask: \emph{Can Bob find Alice's outputs without classical communication?} In what follows, we certainly show that the answer to the question is positive. In this section, we consider Svozil and TB protocols with fewer assumptions, given by replacing classical communications with instantaneous nonlocal effects. Then, we get the same results as the previous ones.

\subsection{Nonlocal description of Svozil's protocol}
Before investigating the nonlocal TB protocol, here, we modify Svozil's argument \cite{svozil} by replacing classical communications with instantaneous
nonlocal effects. For simplicity, we consider a similar notation as in Sec. (II-C). The nonlocal description of Svozil's
protocol proceeds as follows: Parties share independent random
variables $\hat{\lambda}$ and $\hat{\Delta}(\pi/2)$. Alice measures along $\hat{a}$ and her outputs are
$\alpha_{k}=-{\rm sgn}(\hat{a}\cdot\hat{\lambda}_{k})$. Alice's
measurement causes an instantaneous nonlocal effect on Bob's measurement
outputs' so that if Bob measures in the $\hat{b}$
direction, his outputs will be
$\beta_{k}={\rm sgn}[\hat{b}\cdot(\hat{\lambda}_{k}+c_{k}\hat{\Delta}_{k})]$,
where
$c_{k}={\rm sgn}(\hat{a}\cdot\hat{\lambda}_{k}){\rm sgn}(\hat{a}\cdot\hat{\Delta}_{k}))$
(Fig. \ref{3}). Let us select a subset of hidden variables and consider the collection:
\begin{eqnarray}\label{S}
&&\left\{\left(\hat{\lambda}_{k},\hat{\Delta}_{k}\right)\right\},
\end{eqnarray}
where, $\hat{\lambda}_{k}\cdot\hat{\Delta}_{k}=0$, $\hspace{.1cm}\hat{\lambda}_{k+1}=R_{clockwise}(\delta)\hat{\lambda}_{k},\hspace{.2cm} \hat{\Delta}_{k+1}= R_{clockwise}(\delta)\hat{\Delta}_{k},\hspace{.2cm}\hat{\lambda}_{k,\pm}=
\hat{\lambda}_{k}\pm\hat{\Delta}_{k}, \hspace{.2cm}\forall k,\hspace{.2cm} \hat{\lambda}_{k,+}\cdot\hat{\lambda}_{k,-}=0,$ and $\delta=\frac{\pi}{N}\ll1,\hspace{.1cm}k=0,...,N$. The random variables $\hat{\lambda}_{k,\pm}$ divide the Poincare´ sphere
into four equal quadrants.

\emph{Remark 4.}-- Taking into account the definition of $c_{k}$ and selected
random variables, we know that if $\hat{b}$ lies in the
$(\hat{\lambda}_{k,+},\hat{\lambda}_{k,-})$ or
$(-\hat{\lambda}_{k,+},-\hat{\lambda}_{k,-})$ intervals, Bob cannot deduce the nonlocal effect of $c_{k}$. Yet, for other ranges, he
can exactly attain the amount of $c_{k}$ by:
\begin{eqnarray}\label{Non3}
&&\hat{b}\in(\hat{\lambda}_{k,+},\hat{\lambda}_{k,-}),\hspace{.2cm}\beta_{k}=\pm1\rightarrow
c_{k}=?,\\
&&\hat{b}\in(-\hat{\lambda}_{k,+},-\hat{\lambda}_{k,-}),\hspace{.2cm}\beta_{k}=\pm1\rightarrow
c_{k}=?,\\
&&\hat{b}\in(\hat{\lambda}_{k,+},-\hat{\lambda}_{k,-}),\hspace{.2cm}\beta_{k}=\pm1\rightarrow
c_{k}=\pm1,\\
&&\hat{b}\in(-\hat{\lambda}_{k,+},\hat{\lambda}_{k,-}),\hspace{.2cm}\beta_{k}=\pm1\rightarrow
c_{k}=\mp1,\hspace{.1cm}\forall k.
\end{eqnarray}

The collection (\ref{S}) assures Bob that if $\hat{b}$ lies in one of (8) or (9) intervals, after some round of experiment the sign of Bob's outputs will switch to negative values as are given by the following quantities:
\begin{eqnarray*}
..., \beta_{l-1}(\pi/2,
\hat{\lambda}_{l-1})=+1, \beta_{l}(\pi/2,
\hat{\lambda}_{l})=+1,\\ \beta_{l+1}(\pi/2,
\hat{\lambda}_{l+1})=-1, \beta_{l+2}(\pi/2,
\hat{\lambda}_{l+2})=-1,....
\end{eqnarray*}
Here, we assumed that in the $l+1$-th round of the protocol the sign of Bob's outputs has changed and so according
to \emph{Remark 1}, Bob deduces sequences of nonlocal effects $c_{k}$ as the following:
\begin{eqnarray}\label{Sec1}
...,c_{(l-1)}=+1,c_{l}=+1,c_{(l+1)}=-1,c_{(l+2)}=-1,...,
\end{eqnarray}
Therefore, Bob concludes that $\hat{a}$ is in the same (or in the opposite) direction of the unit vector $\hat{\lambda}_{l}$, with the uncertainty factor $\delta$. At this stage, if Alice informs Bob of only one of her outputs, for example,
$\alpha(\hat{a},\lambda_{r})=-{\rm sgn}(\hat{a}.\hat{\lambda}_{r})=+1$
(or $-1$), Bob will infer that $\hat{a}$ lies in the down (up) semicircle \cite{Note1}. Based on what we have shown, the reader can admit that Bob can deduce Alice's measurement setting by considering every two subsets of SRV.

The study of two special cases of Alice's measurement settings seems interesting.

\emph{Remark 5.}-- If the angle between measurement
settings of the parties is equal to $\mid\varphi_{\hat{a}}-\varphi_{\hat{b}}\mid=\pi/4$ or $3\pi/4$ then Bob will get
one of the following outputs:
\begin{eqnarray}\label{Non4}
\text{If}\left\{
  \begin{array}{ll}
    \beta_{k}=+1\hspace{.1cm}\text{for}\hspace{.1cm}
\hat{b}\in(-\hat{\lambda}_{k,+},+\hat{\lambda}_{k,-}),\text{and} \\
   \beta_{k}=-1
\hspace{.1cm}\text{for}\hspace{.1cm}
\hat{b}\in(+\hat{\lambda}_{k,+},-\hat{\lambda}_{k,-}),\hspace{.5cm}\forall
k,
  \end{array}
\right\}
\text{then, Bob will deduce}\hspace{.1cm}\mid\varphi_{\hat{a}}-\varphi_{\hat{b}}\mid=\pi/4\hspace{.2cm}\text{or}\hspace{.2cm}
3\pi/4,
\hspace{.1cm}\Rightarrow\hat{a}=\pm
R_{clockwise}(\pi/4)\hat{b},\nonumber
\end{eqnarray}
where Bob obtains Alice's measurement direction ($\pm\hat{a}$) by
rotating $\hat{b}$ clockwise around the center by the value of
$\pi/4$.
\begin{eqnarray}\label{Non5}
\text{If}
\left\{
  \begin{array}{ll}
   \beta_{k}=-1\hspace{.1cm}\text{for}\hspace{.1cm}
\hat{b}\in(-\hat{\lambda}_{k,+},+\hat{\lambda}_{k,-}),\text{and} \\
   \beta_{k}=+1
\hspace{.1cm}\text{for}\hspace{.1cm}
\hat{b}\in(+\hat{\lambda}_{k,+},\hat{-\lambda}_{k,-}),\hspace{.5cm}\forall
k,
  \end{array}
\right\}
\text{then, Bob will deduce}\hspace{.1cm}\mid\varphi_{\hat{a}}-\varphi_{\hat{b}}\mid=\pi/4
\hspace{.1cm}\text{or}\hspace{.2cm} 3\pi/4,
\hspace{.1cm} \Rightarrow\hat{a}=\pm
R_{c-clockwise}(\pi/4)\hat{b},\nonumber
\end{eqnarray}
where Bob obtains $\pm\hat{a}$ by
rotating $\hat{b}$ counterclockwise around the center by the value of $\pi/4$. Therefore, if Alice informs Bob of only one of her outputs, he will exactly deduce the $\hat{a}$ direction without any need for further information.

\subsection{Nonlocal description of TB model}
\begin{figure}
\centering
\includegraphics[height=4cm,width=8cm]{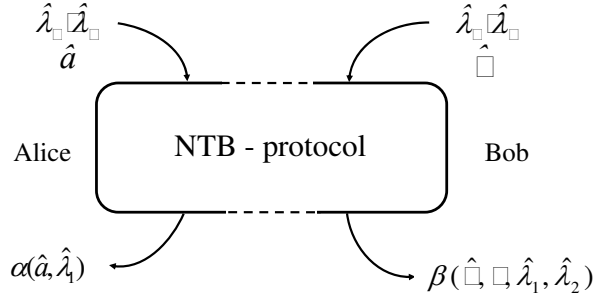}
\caption{NTB protocol as an elementary resource for simulating
physical systems, where parties shared two sets of independent
random variables $\hat{\lambda}_{1}$ and $\hat{\lambda}_{2}$;
Alice and Bob inputs are $\hat{a}$ and $\hat{b}$, respectively.
NTB protocol outputs are $\alpha=-{\rm sgn}(\hat{a}.\hat{\lambda}_{1})$
and $\beta={\rm sgn}[\hat{b}.(\hat{\lambda}_{1}+c\hat{\lambda}_{2})]$,
where
$c={\rm sgn}(\hat{a}\cdot\hat{\lambda}_{1}){\rm sgn}(\hat{a}\cdot\hat{\lambda}_{2})$.}
\label{2}
\end{figure}

In this subsection, we suggest a nonlocal version of the TB protocol (NTB) which is an imaginary device that includes two input-output ports, one at Alice's location and another at Bob's, while Alice and Bob are spacelike separated. NTB protocol proceeds as follows: The parties share two independent random variables $\hat{\lambda}_{1}$ and $\hat{\lambda}_{2}$. Alice measures along $\hat{a}$ and her output is $\alpha=-{\rm sgn}(\hat{a}\cdot\hat{\lambda}_{1})$. Alice's measurement causes a nonlocal effect on Bob's measurement outputs' so that if Bob's measurement setting is in the $\hat{b}$ direction, his output is $\beta={\rm sgn}[\hat{b}\cdot(\hat{\lambda}_{1}+c\hat{\lambda}_{2})]$, where $c={\rm sgn}(\hat{a}\cdot\hat{\lambda}_{1}){\rm sgn}(\hat{a}\cdot\hat{\lambda}_{2})$ (Fig. \ref{2}).

\emph{Remark 6.}-- The TB and the NTB frameworks are equivalent at the level of what they aim to calculate; we can replace one bit classical communication in the Toner and Bacon model \cite{Bacon} with a nonlocal effect so that the marginal and joint probabilities calculated in either of these scenarios are similar to those within the other one. In the NTB model, Alice and Bob know directions of random variables $\hat{\lambda}_{1}$ and $\hat{\lambda}_{2}$ for each round of the protocol, but, the values of $c$ are not accessible to Bob.

Similar to the Sec. (II-A), we consider the unit vectors $\hat{\lambda}_{1}$ and $\hat{\lambda}_{2}$ in the spherical coordinate ($\theta,\phi$) in the ranges of $\theta\in(0,\pi)$ and $\phi\in(0,2\pi)$, and divide them into $N$ equal parts ($\delta=\pi/N\ll1$, with $N\rightarrow\infty$). To show that selected SRV subsets (Sec. II-A) are not restricted to collection \ref{xy}, we select a subset of SRV in which the elements of each pair are orthogonal.

In the first case, we select a subset of SRV which lies in the $xy$ plane. In the Poincare´ sphere coordinates, the selected SRV is represented by $\left\{\hat{\lambda}^{xy}_{1}(\theta=\pi/2,l\delta),\hat{\lambda}^{xy}_{2}(\theta=\pi/2,k\delta)\right\} \equiv\left\{\hat{\lambda}^{xy}_{1,l},\hat{\lambda}^{xy}_{2,k}\right\}$, where $l,k=0,1,...,N$. Let us select a subset of SRV and consider the collection:
\begin{eqnarray}\label{S2}
\left\{\left(\hat{\lambda}^{xy}_{1,l},\hat{\lambda}_{2,l}^{xy}\right)\right\},
\end{eqnarray}
where, $\hat{\lambda}_{2,l}^{xy}=R(\hat{z},\pi/2)\hat{\lambda}_{1,l}^{xy}$, $\hspace{.2cm}$ $\hat{\lambda}_{1,l}^{xy}\cdot\hat{\lambda}_{2,l}^{xy}=0$, and $\hat{\lambda}_{i,l+1}^{xy}=R(\hat{z},\delta)\hat{\lambda}_{i,l}^{xy}$, $\hspace{.2cm}i=1,2,\hspace{.2cm}l=0,...,N,\hspace{.2cm}\forall
l$. Moreover, we define random variables $\hat{\lambda}_{l,\pm}^{xy}=\hat{\lambda}_{1,l}^{xy}\pm\hat{\lambda}_{2,l}^{xy},$ $\hspace{.2cm}\hat{\lambda}_{l,+}^{xy}\cdot\hat{\lambda}_{l,-}^{xy}=0,\hspace{.2cm}\forall l$.
The random variables $\hat{\lambda}_{l,\pm}$ divide the Poincare´ sphere
into four equal parts. The other elements of set (\ref{S2}) are given
by rotating $\hat{\lambda}_{l,\pm}$ around the $\hat{z}$ axis by the value of $\delta$,
$\hat{\lambda}_{l+1,\pm}^{xy}=R(\hat{z},\delta)\hat{\lambda}_{l,\pm}^{xy}$.
\begin{figure}
\centering
\includegraphics[height=7cm,width=7.5cm]{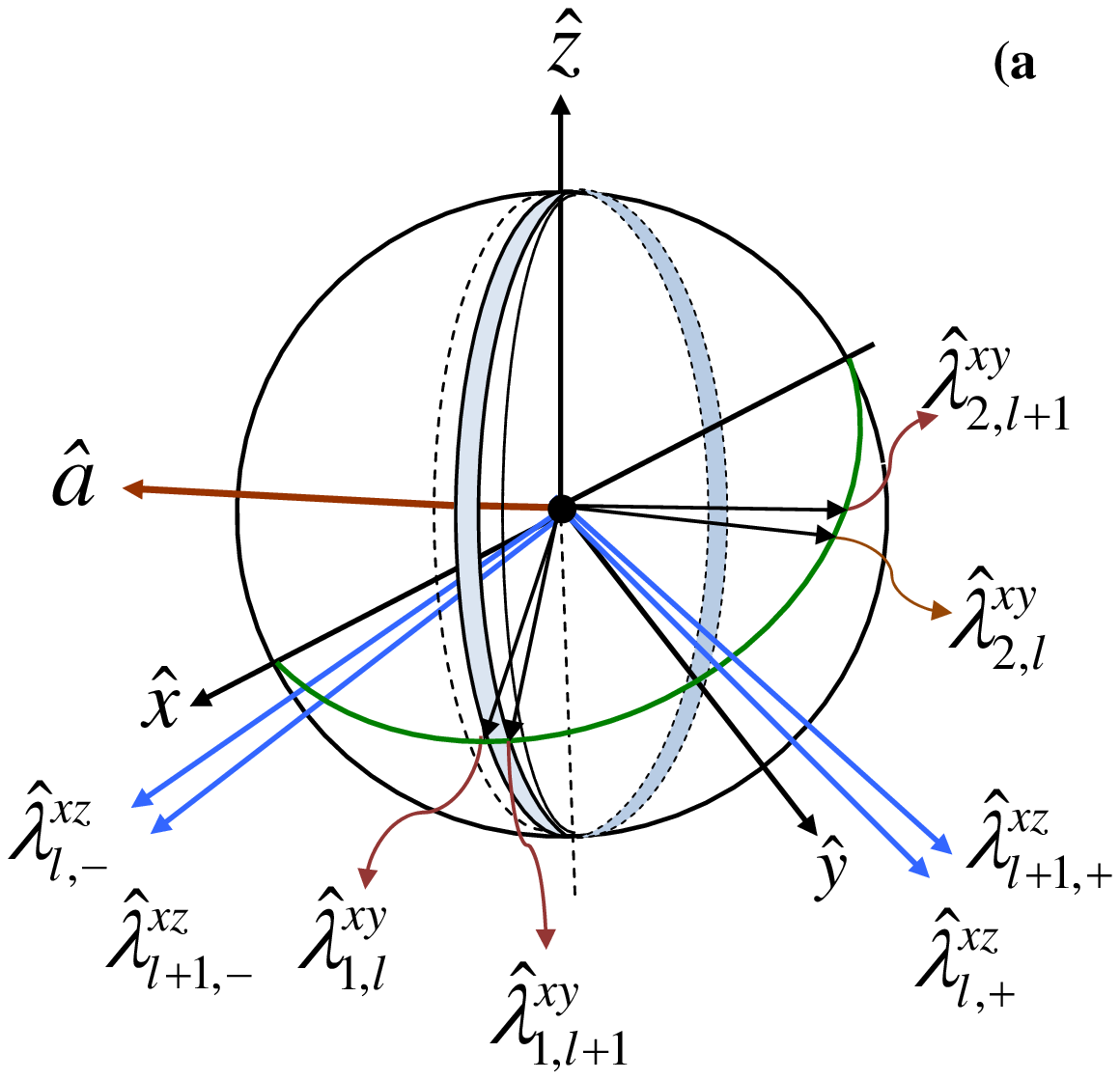}\\
\includegraphics[height=7.2cm,width=7.5cm]{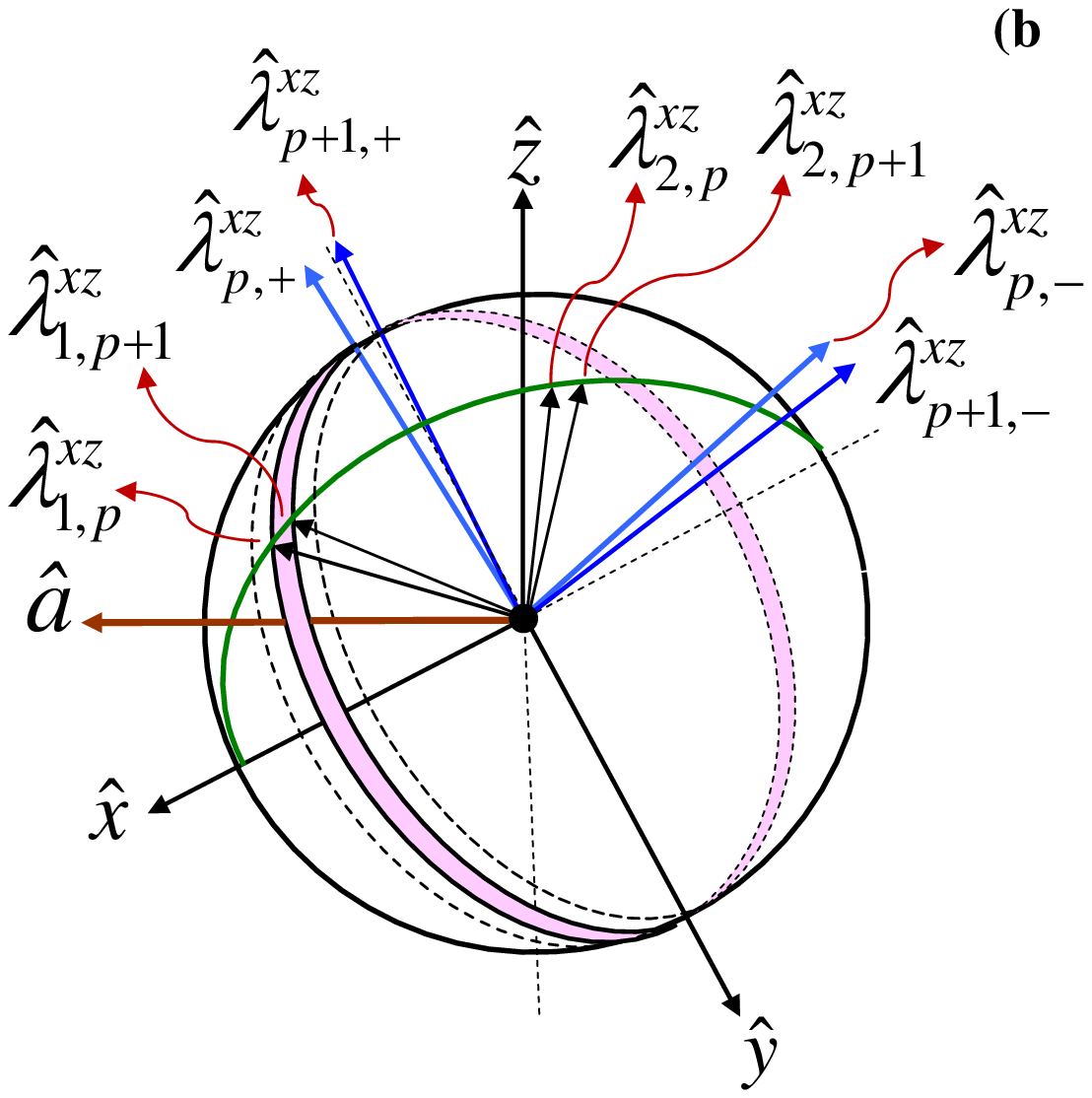}
\caption{(Color online). (a) Subsets of shared random variables lie in the $xy$
plan. The blue zone defines a plane with unit vectors
$\hat{\lambda}_{2,l}^{xy}$. According to the definition of $c$,
Alice's measurement setting $\hat{a}$ must lie in the blue zone.
(b) Subsets of shared random variables lie in the $xz$ plane. The
red zone defines a plane with unit vectors
$\hat{\lambda}_{2,p}^{xz}$, so that Alice's measurement setting
$\hat{a}$ must lie in the red zone. In fact, these thin strips
sweep the surface of the unit sphere.}\label{4}
\end{figure}

\emph{Remark 7.}-- Concerning the definition of $c_{l}$ and the selected random variables in (\ref{Non3})-(9), we know that if $\hat{b}$ lies in the
$(\hat{\lambda}^{xy}_{l,+},\hat{\lambda}^{xy}_{l,-})$ or
$(-\hat{\lambda}^{xy}_{l,+},-\hat{\lambda}^{xy}_{l,-})$ intervals,
Bob cannot deduce the nonlocal effect of $c_{l}$, but for other
ranges, he can exactly attain the value of $c_{l}$.

The collection (\ref{S2}) assures Bob that if $\hat{b}$ lies in one of the (8) or (9) intervals,
his corresponding outputs will satisfy the following sequences:
\begin{eqnarray}\label{beta}
&&...,\beta_{l-1}(c_{l-1},\hat{\lambda}_{1,l-1}^{xy},\hat{\lambda}_{2,l-1}^{xy})=+1,
\beta_{l}(c_{l},\hat{\lambda}_{1,l}^{xy},\hat{\lambda}_{2,l}^{xy})=+1,\nonumber\\
&&\beta_{l+1}(c_{l+1},\hat{\lambda}_{1,l+1}^{xy},\hat{\lambda}_{2,l+1}^{xy})=-1,
\beta_{l+2}(c_{l+2},\hat{\lambda}_{1,l+2}^{xy},\hat{\lambda}_{2,l+2}^{xy})
=-1,....\hspace{.4cm}
\end{eqnarray}

Similar to the previous case, we assume here that in the $l+1$-th round of the protocol, the sign of Bob's outputs has changed and similar to \emph{Remark 3}, Bob deduces sequences of nonlocal effects of $c_{k}$ as following:
\begin{eqnarray}\label{Sec1}
..., c^{xy}_{(l-1)}=+1, c^{xy}_{l}=+1, c^{xy}_{(l+1)}=-1, c^{xy}_{(l+2)}=-1,....\nonumber\\
\end{eqnarray}
Therefore, Bob infers that $\hat{a}$ lies in the plane with unit vector
$\hat{\lambda}^{xy}_{2,l}$, with the uncertainty factor $\delta$ [Fig.
\ref{4}(a)] \cite{Note2}. In fact, $\hat{a}$ located in the uncommon parts of two semispheres which are defined by $\hat{\lambda}^{xy}_{2,l}$ and
$\hat{\lambda}^{xy}_{2,l+1}$ [Fig. \ref{4}(a)].

In the next step of protocol, parties select another subset of SRV in the $xz$
plane as:
\begin{eqnarray}\label{S11}
\left\{\left(\hat{\lambda}^{xz}_{1,p},\hat{\lambda}_{2,p}^{xz}\right)\right\},
\end{eqnarray}
where, $\hat{\lambda}_{1,p}^{xz}\cdot\hat{\lambda}_{2,p}^{xz}=0$, $\hspace{.2cm}\hat{\lambda}_{2,p}^{xz}=R(\hat{y},\pi/2)\hat{\lambda}_{1,p}^{xz}$,   $\hspace{.2cm}\hat{\lambda}_{j,p+1}^{xz}=R(\hat{y},\delta)\hat{\lambda}_{j,p}^{xz}$ $\hspace{.2cm}\forall
p$, and $\hspace{.2cm}\delta=\frac{\pi}{N}\ll1,\hspace{.2cm}j=1,2,\hspace{.2cm}p=0,...,N$. Moreover, we define random variables $\hat{\lambda}_{p,\pm}^{xz}=\hat{\lambda}_{1,p}^{xz}\pm\hat{\lambda}_{2,p}^{xz},$ $\hspace{.2cm}\hat{\lambda}_{p,+}^{xz}\cdot\hat{\lambda}_{p,-}^{xz}=0,\hspace{.2cm}\forall p$. The random variables $\hat{\lambda}_{p,\pm}^{xz}$ divide the Poincare´ sphere into four equal parts. The other elements of set (\ref{S11}) are given by rotating $\hat{\lambda}_{p,\pm}$ around the $\hat{y}$ axis by the value of $\delta$,
$\hat{\lambda}_{p+1,\pm}^{xz}=R(\hat{y},\delta)\hat{\lambda}_{p,\pm}^{xz}$.

Bob's outputs are similar to (\ref{beta}), with
$\beta_{l}(c_{l},\hat{\lambda}_{1,l}^{xy},\hat{\lambda}_{2,l}^{xy})
\rightarrow\beta_{p}(c_{p},\hat{\lambda}_{1,p}^{xz},\hat{\lambda}_{2,p}^{xz})$
and $c^{xy}_{l}\rightarrow c^{xz}_{p}$.
With this sequence, Bob infers that $\hat{a}$ lies in the plane with unit vector
$\hat{\lambda}^{xz}_{2,p}$, with the uncertainty factor $\delta$ [Fig.
\ref{4}(b)].

The subsets (\ref{S2}) and (\ref{S11}) define two strips as Figs. \ref{4}(a) and \ref{4}(b) show. These two strips cross each other at two points [similar to Fig. \ref{1}(c)]. Alice's measurement setting $\hat{a}$ is in the same (or in the opposite) direction as the unit vector that connects the origin of the Poincare´ sphere to the
cross points [Fig. \ref{1}(c)]. Similar to what happens in the previous case, if Alice informs Bob of only one of her outputs, he exactly deduces the $\hat{a}$ direction without any need for further information.

Here, we discuss two interesting cases in Alice's measurements setting.

\emph{Remark 8.}-- If the angle between the measurement
settings of parties in the $xy$ plane is equal to $\mid
\phi_{\hat{a}}-\phi_{\hat{b}} \mid=\pi/4$ or $3\pi/4$, Bob's
measurement outputs are given as the following: \\
\begin{eqnarray*}\label{Non14}
&&\text{If}
\left\{
  \begin{array}{ll}
   \beta_{k}=+1\hspace{.1cm}\text{for}\hspace{.1cm}
\hat{b}\in(-\hat{\lambda}^{xy}_{k,+},+\hat{\lambda}^{xy}_{k,-}),\text{and} \\
 \beta_{k}=-1
\hspace{.1cm}\text{for}\hspace{.1cm}
\hat{b}\in(+\hat{\lambda}^{xy}_{k,+},\hat{-\lambda}^{xy}_{k,-}),\hspace{.5cm}\forall
k
  \end{array}
\right\}
\hspace{.2cm}\text{then, Bob will deduce}
\hspace{.1cm}\mid\phi_{\hat{a}}-\phi_{\hat{b}}\mid=\pi/4\hspace{.1cm}
\text{or}\hspace{.1cm} 3\pi/4,\nonumber\\
&&\hspace{9cm}\Rightarrow \hat{a} \hspace{.1cm}\text{lies in the
plane with unit vector}\hspace{.2cm}\hat{n}=+
R_{\hat{\phi}}(\hat{z},\pi/4)\hat{b},\nonumber
\end{eqnarray*}
where Alice's measurement setting ($\hat{a}$) lies in a plane with
the unit vector $\hat{n}$ by
rotating $\hat{b}$ around the $\hat{z}$ axis (in the $+\hat{\phi}$
direction) by $+\pi/4$.
\begin{eqnarray*}\label{Non14}
&&\text{If}
\left\{
  \begin{array}{ll}
   \beta_{k}=-1\hspace{.1cm}\text{for}\hspace{.1cm}
\hat{b}\in(-\hat{\lambda}^{xy}_{k,+},+\hat{\lambda}^{xy}_{k,-}), \text{and}\\
   \beta_{k}=+1
\hspace{.1cm}\text{for}\hspace{.1cm}
\hat{b}\in(+\hat{\lambda}^{xy}_{k,+},\hat{-\lambda}^{xy}_{k,-}),\hspace{.5cm}\forall
k
  \end{array}
\right\}\hspace{.2cm}\text{then, Bob will deduce}\hspace{.1cm}\mid\phi_{\hat{a}}-\phi_{\hat{b}}\mid=\pi/4\hspace{.1cm}
\text{or}\hspace{.1cm} 3\pi/4,\\
&&\hspace{9cm}\Rightarrow \hat{a} \hspace{.1cm}\text{lies in the plane
with unit vector}\hspace{.2cm}\hat{n}=+
R_{\hat{\phi}}(\hat{z},-\pi/4)\hat{b},\hspace{.3cm}\nonumber
\end{eqnarray*}
where Alice's measurement setting ($\hat{a}$) lies in a plane with
the unit vector $\hat{n}$ by
rotating $\hat{b}$ around the $\hat{z}$ axis (in the $+\hat{\phi}$
direction) by $-\pi/4$. Similar to the above approach, if the angle between measurement
settings of parties in the $xz$ plane is equal to $\pi/4$ or $3\pi/4$, $\hat{a}$ lies in a plane with the
unit vector $\hat{n'}$, and Bob obtains the direction $\hat{n'}$ by
rotating $\hat{b}$ around the $\hat{y}$ axis by $\pm\pi/4$.

\section{Summery and outlooks}

In this paper, we reviewed TB and Svozil protocols and showed that
if parties select two subsets of SRV, Bob can
deduce all of Alice's measurement.

\begin{eqnarray}
  \begin{array}{ll}
    \text{\textbf{The original TB protocol} (SRV + 1 cbit for each round)} \\
$\hspace{1cm}$\text{
+ \textbf{Alice informs Bob of one of her outputs} }
  \end{array}
\longrightarrow
  \begin{array}{ll}
    \text{\textbf{Bob deduces Alice's measurement} } \\
$\hspace{1cm}$\text{\textbf{outputs with certainty}}\hspace{.1cm} \delta\nonumber
  \end{array}
\end{eqnarray}

Afterwards, we suggested a nonlocal version of TB and Svozil protocols by replacing classical communications with nonlocal effects and obtained the same results as mentioned in the previous part.

\begin{eqnarray}
  \begin{array}{ll}
    \text{\textbf{The nonlocal version of TB protocol} (SRV  + Alice's measurement } \\
$\hspace{3cm}$\text{causes a nonlocal effect on Bob's output)} \\
$\hspace{1.3cm}$\text{
+ \textbf{Alice informs Bob from one of her outputs} }
  \end{array}
\longrightarrow
  \begin{array}{ll}
    \text{\textbf{Bob deduces Alice's measurement} } \\
$\hspace{1cm}$\text{\textbf{outputs with certainty}}\hspace{.1cm} \delta\nonumber
  \end{array}
\end{eqnarray}

Here, a question arises: \emph{are TB and NTB protocols causal?} In the NLB-box \cite{PR1}, if Alice's (Bob's) input is $x=0$ ($y=0$), she (he) can distinguish the other one's output exactly. Otherwise, she (he) doesn't have any information about his (her) outputs. It is usually interpreted that the NLB-box is causal. We know that the NTB protocol cannot be used for Superluminal signaling in each rounds of the protocol. Only, after the $l+p$ rounds of the protocol, Bob concludes that
$\hat{a}$ is orthogonal to $\hat{\lambda}^{xy}_{2,l}$ and
$\hat{\lambda}^{xz}_{2,p}$ directions. He must await for Alice's message. Only at this stage, he can know Alice's measurement setting exactly. We know that NLB represents undirected resources, but NTB represents directed ones that can be shared between two parties \cite{BP}. Hence, in the NLB approach, Bob does not have complete information about
Alice's inputs. Yet, in our description Bob gets complete information
about each of Alice's results.

As we know, Cerf \emph{et al.} \cite{Cerf} suggested a kind of NLB-box based on the TB protocol which perfectly simulated a maximally entangled (singlet) state by using one instance of the NLB-box machine and no communication at all. The NTB protocol can be used for discussing the communication complexity problem. In this approach, Alice and Bob shared an NTB machine as well as
shared random variables in the form of the pairs of normalized vectors $\hat{\lambda}_{1}$ and $\hat{\lambda}_{2}$, randomly and independently distributed over the Poincare´ sphere. $n$-tuple of inputs is denoted as $\hat{a}\equiv(x_{1},x_{2},...,x_{n})$ and $\hat{b}\equiv(y_{1},y_{2},...,y_{n})$ are the vectors that determine Alice and Bob measurements, respectively (where, $x_{i},y_{j}\in\{0,1\}$, and $i,j=0,1,...,n$). With due attention to our approach, we proved that the availability of perfect the NTB protocol makes the communication complexity of all Boolean functions trivial. Therefore, TB's claim that Bob obtains ``no information" about Alice's outputs from the classical communications, is not correct. It seems that the TB protocol used some unacceptable concepts in its approach, and consequently the question about ``what minimum classical resources are required to simulate quantum correlations?" is still open \cite{Comm}.

Moreover, in TB and NTB protocols, the parties have unrestricted control of the SRV. Therefore, here is another interesting model in which the parties have partial information (or don't have any information) about the SRV. It sheds light on quantum-entanglement notation.

$\vspace{.1cm}$

\textbf{Acknowledgments:} We thank E. Azadeghan for discussions and reading our manuscript.





\begin{references}

\bibitem {EPR} A. Einstein, B. Podolsky, and N. Rosen, Can quantum-mechanical description of physical reality be considered complete? Phys. Rev. \textbf{47}, 777-780 (1935).

\bibitem {Bell} J. S. Bell, On the Einstein, Podolsky, Rosen paradox, physics (Long Island City, N.Y.) \textbf{1},
195 (1964).

\bibitem {Bell1} J. S. Bell, \emph{Speakable and Unspeakable in
Quantum Mechanics} (Cambridge Univ. Press, Cambridge, U.K.)
(1993).

\bibitem{CHSH} J. F. Clauser, M.A. Horne, A. Shimony, and R. A. Holt, Proposed experiment to test local hidden-variable theories, Phys. Rev. Lett. \textbf{23}, 880 (1969).

\bibitem{CH} J.F. Clauser, and M.A. Horne, Experimental consequences of objective local theories, Phys.
Rev.D $\mathbf{10}$, 526 (1974).




\bibitem{As} A. Aspect, P. Grangier, and G. Roger, Experimental realization of Einstein-Podolsky-Rosen-Bohm gedankenexperiment: A new violation of Bell's inequalities,
Phys. Rev. Lett. \textbf{49}, 91-94 (1982);
W. Tittel, J. Brendel, H. Zbinden, and N. Gisin, Violation of Bell inequalities by photons more than 10 km apart, Phys. Rev. Lett. \textbf{81}, 3563-3566 (1998);
J. Pan, \emph{et al.}, Experimental test of quantum nonlocality in three-photon Greenberger, Horne, Zeilinger
entanglement, Nature \textbf{403}, 515 (2000);
M. A. Rowe, \emph{et al.}, Experimental violation of a Bell's inequality with efficient detection, Nature \textbf{409}, 791 (2001);
C. A. Sackett, \emph{et al.}, Experimental entanglement of four particles, Nature \textbf{404}, 256 (2000).








\bibitem {Mau} T. Maudlin, in \emph{PSA 1992, Volume 1}, edited by D. Hull, M. Forbes,
and K. Okruhlik (Philosophy of Science Association, East Lansing,
1992), pp. 404--417.






\bibitem{Brass} G. Brassard, R. Cleve, and A. Tapp, Cost of exactly simulating quantum entanglement with classical communication, Phys. Rev. Lett. \textbf{83}, 1874 (1999); G. Brassard, Quantum communication complexity, Found. Phys. \textbf{33}, 1593 (2003)
(quant-ph/0101005).

\bibitem{St} M. Steiner, Towards quantifying nonlocal information transfer: finite-bit
nonlocality, Phys. Lett. A \textbf{270}, 239
(2000).

\bibitem{Gisin1} B. Gisin and N. Gisin, A local hidden variable model of quantum correlation exploiting
the detection loophole, Phys. Lett. A 260, 323 (1999).


\bibitem{Cerf1} N. J. Cerf, N. Gisin, and S. Massar, Classical teleportation of a quantum bit, Phys. Rev. Lett. 84,
2521 (2000).


\bibitem{Bacon} B. F. Toner, and D. Bacon, Communication cost of simulating Bell correlations, Phys. Rev. Lett. \textbf{91}, 187904 (2003).


\bibitem {svozil} K. Svozil, Communication cost of breaking the Bell barrier, Phys. Rev. A \textbf{72}, 050302(R) (2005); Erratum: Communication cost of breaking the Bell barrier, \emph{ibid} \textbf{75}, 069902(E)(2007).



\bibitem{Cav1} T. E. Tessier, C. M. Caves, I. H. Deutsch, B. Eastin, and D. Bacon, Optimal classical-communication-assisted local model of $n$-qubit Greenberger-Horne-Zeilinger correlations, Phys. Rev. A \textbf{72}, 032305
(2005).

\bibitem{Cav2} J. Barrett, C. M. Caves, B. Eastin, M. B. Elliott, and S. Pironio, Modeling Pauli measurements on graph states with nearest-neighbor classical communication, Phys. Rev. A \textbf{75}, 012103
(2007).




\bibitem{De} J. Degorre, S. Laplante and J. Roland, Classical simulation of traceless binary observables on any bipartite quantum state, Phys. Rev. A \textbf{75}, 012309 (2007); Simulating quantum correlations as a distributed sampling problem, \emph{ibid} \textbf{72},
062314 (2005).




\bibitem{Non} S. Gr\"{o}blacher, \emph{et al.}, An experimental test of nonlocal realism, (Supplementary information part I), Nature \textbf{446}, 871,
(2007).


\bibitem{PR1} S. Popescu, and D. Rohrlich, Quantum nonlocality as an axiom, Found. Phys. \textbf{24}, 379 (1994).







\bibitem{JM} N. S. Jones and Ll. Masanes, Interconversion of nonlocal correlations, Phys. Rev. A
\textbf{72}, 052312, (2005).



\bibitem{Bra2} G. Brassard, H. Buhrman, N. Linden, A. A. M\'{e}thot, A. Tapp, and F. Unger, Limit on nonlocality in any World in which communication complexity is not trivial, Phys. Rev. Lett. \textbf{96}, 250401
(2006).



\bibitem{Dam} W. van Dam, PhD thesis, Univ. Oxford (2000); quant-ph/0501159.



\bibitem{IC} M. Pawlowski \emph{et al.}, Information causality as a physical principle, Nature \textbf{461}, 1101, (2009).





\bibitem{BGS} N. Brunner, N. Gisin and V. Scarani, Entanglement and nonlocality are different resources, New Journal of Physics \textbf{7} 88 (2005).















\bibitem{Cerf} N. J. Cerf, N. Gisin, S. Massar, and S. Popescu, Simulating maximal quantum entanglement without communication, Phys. Rev. Lett.
\textbf{94}, 220403 (2005).














\bibitem{Note1} If classical communicating bits are given by reverse sign
$...,c_{(l-1)}=-1,c_{l}=-1,c_{(l+1)}=+1,c_{(l+2)}=+1,...$, then, either
Alice's measurement setting $\hat{a}$ will be orthogonal to
$\hat{\lambda}_{l+1}$ or $\hat{a}$ will be in the same (or in opposite) direction of the unit vector
 $\hat{\Delta}_{l+1}$ (with uncertainty factor $\delta$).

\bibitem{Note2} If nonlocal effects $c_{k}$ are given by reverse sign
$..., c^{xy}_{(l-1)}=-1, c^{xy}_{l}=-1, c^{xy}_{(l+1)}=+1,
c^{xy}_{(l+2)}=+1,...,$, Bob will deduce that $\hat{a}$ is orthogonal
to the $\hat{\lambda}^{xy}_{1,l+1}$ direction (with uncertainty factor $\delta$).


\bibitem{BP} J. Barrett, and S. Pironio, Popescu-Rohrlich Correlations as a unit of nonlocality, Phys. Rev. Lett. \textbf{95}, 140401 (2005).



\bibitem{Comm} In another work, we try to clarify this point.



























































\end{references}
\end{document}